# POLARIZATION OF THE TWILIGHT SKY: MEASUREMENTS, EMPIRICAL MODEL, RADIATIVE TRANSFER SIMULATION AND APPLICATION TO AEROSOL PROFILE RETRIEVAL

*O.V. Postylyakov*[1,2], *O.S. Ugolnikov*[3,4], *I.A. Maslov*[3], *A.N. Masleev*[1]

[1]A.M. Obukhov Institute of Atmospheric Physics, Moscow, Russia
[2]Finnish Meteorological Institute, Helsinki, Finland
[3]Space Research Institute, Moscow, Russia
[4]Astro-Space Center, P.N. Lebedev Physical Institute, Moscow, Russia

*ovp@omega.ifaran.ru, ugol@tanatos.asc.rssi.ru*

## 1. INTRODUCTION

Polarization observations offer new opportunities in atmosphere aerosol sounding, and the POLDER space instrument and the AERONET ground network began to apply it. Using the observations of the degree of linear polarization they determine aerosol characteristics averaged over column with predomination of the lower-troposphere part of column. We investigate the opportunity to determine the vertical profile of aerosol characteristics using polarization observations during twilight period.

The essential trouble in solution of the raising aerosol inverse problem is multiple scattering of light, which may have the same depolarizing effect like the aerosol. We had proposed the algorithm for retrieval of the aerosol characteristic vertical profile and carried out the error analysis in approximation of single scattering (Section 6.3). To evaluate the fraction of multiple scattering we had developed both its semi-empirical model, basing on observations (Section 3), and a radiative transfer model, which take into account all orders of scattering (Section 4). The main goal of this paper is a comparison of polarized radiance and characteristics derived from the polarization measurements by semi-empirical model with the same characteristics derived from numerical simulation of the radiative transfer for different aerosol (Sections 6.1, 6.2).

Definitions of used terms are given in Appendix A.

## 2. INSTRUMENTATION AND OBSERVATIONS

First session of polarimetric observations was carried out in summer 1997 at the Astronomical Observatory of Odessa University, Ukraine (46.40°N, 30.25°E). The solar vertical was scanned in a range of zenith distances $z$ from –70° to +70°, the wavelength was equal to 356 nm [1]. Second session of observations was carried out in summer 2000 in South Laboratory of Sternberg Astronomical Institute, Crimea, Ukraine (44.73°N, 34.02°E) [2]. We made the wide-angle polarimetric CCD-images of the near-zenith region with the values of zenith distance from –15° to +15° in four wide colour regions (360, 440, 550 and 700 nm). One more session of 550 nm observations was held at the same place in late 2002. Simultaneous observations at Kislovodsk City and Kislovodsk High-Mountain (43.73°N, 42.66°E, 2070 m.a.s.l.) Stations of the Institute of Atmospheric Physics, at Zvenigorod (55.7°N, 36.8°E) and at Lovozero (67.97°N, 35.02°E) were carried out in 1999-2000 by photometer with rotating polarizing glass. The observations were performed at 800 nm in the zenith direction [3].

## 3. SEMI-EMPIRICAL MODEL FOR MULTIPLE SCATTERING SEPARATION

A method of single and multiple scattered light separation using observations of polarization of twilight sky was developed in [1, 2]. The simplified idea of the method follows. When the Sun is situated near the horizon, the maximum polarization point of single scattering should be situated about 90° from it (or exactly 90° if the scattering is molecular) or near the zenith. And as the Sun is depressing, this point follows it by the vertical to the glow segment, remaining at the constant angular distance from the Sun (since the effective scattering altitude is not changing rapidly at this twilight period). The maximum polarization point for multiple scattering at the sunset is expected to be near the zenith too, but as the secondary light source is the whole sky and basically, the glow area, and not directly related with the Sun, this point should be practically immovable remaining near the zenith. And thus, the position of the maximum polarization point for the total sky background will be the reflection of the single and multiple scattering balance.

A mathematical procedure of the single scattering contribution calculation is described in [1]. This procedure uses the polarization of light incoming from a few zenith directions for a few solar depth angles during twilight. The main quantity used in this method is polarization ratio mixed derivative $d^2K/dzdz_S$ (see Appendix A for definitions). This value strongly depends on the single scattering contribution and aerosol phase matrix at the scattering angle near to 90°, and we can use both sky polarization and this value measured during the observation as the test of different aerosol models.

Figure 1 shows the dependence of the maximum polarization point zenith distance on the solar zenith angle for one date of simultaneous observations in 550 and 700 nm. The blue line is corresponding to the case of pure Rayleigh single scattering. We can see that at light phase of twilight until the Sun depression about 4-5 degrees, the maximum polarization point really follows the Sun, but the velocity of such motion is lower a bit than for the case of pure single scattering. It is pointing to almost constant ratio of single and multiple scattering at this time. And then this point is turning back to the zenith, and polarization profile of the whole solar vertical becomes symmetric relatively the zenith. It can be related only with the single scattering light disappearing on the background of multiple scattering.

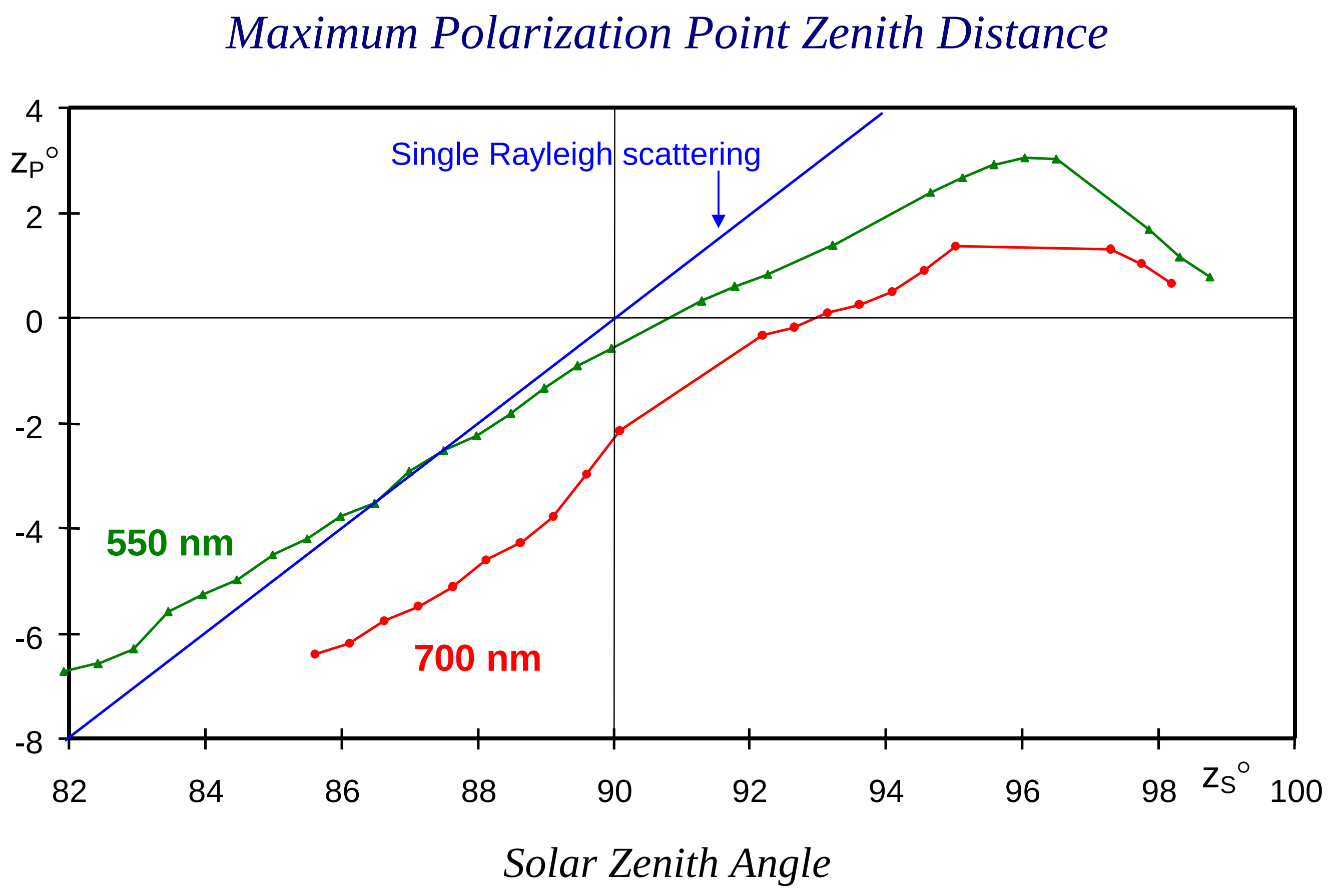

**Figure 1.** The dependency of maximum polarization point zenith distance, $z_P$, on the sun zenith distance, $z_S$, for the evening of August 6th, 2000, 550 and 700 nm.

## 4. RADIATIVE TRANSFER MODEL MCC++

A simulation of the twilight polarized radiance was performed using a Monte Carlo linearized radiative transfer model MCC++ [4]. The MCC++ model calculates radiative transfer in the spherical shell atmosphere with account of all orders of scattering, aerosol and Rayleigh scattering, aerosol and gaseous absorption and surface albedo. The MCC++ makes quick computations: the computation time for 1% accuracy is comparable with pseudospherical radiative transfer models of other authors. The code was validated against other radiative transfer models for the twilight observations [5]. For usage in retrieval problems the model may perform quick calculation of weighting functions [6].

Calculations were carried out for the solar zenith angle up to 96° at wavelengths equal to 360, 440, 550, 700, 800 nm.

## 5. MODELS OF ATMOSPHERIC AEROSOL

A few types of the aerosol microphysical characteristics proposed by WMO (see [7] and Appendix C) were used for radiative transfer calculations, simulating the different aerosol scenarios. The urban, continental and maritime types of aerosol from 0 to 10 km were compared with observations. The stratospheric type of aerosol at latitudes above 10 km was used in all cases. The phase matrix, absorption and scattering cross sections of aerosol were calculated using Mie theory.

MODTRAN aerosol extinction profile at 600 nm, corresponding to surface visible range 50 km and background stratospheric aerosol, was used in all cases.

The other atmospheric characteristics are given in Appendix B.

## 6. ANALYSIS

### 6.1. Degree of linear polarization

Figure 2a shows observed degree of polarization in the zenith against calculated by the radiative transfer model for the urban type of aerosol and background aerosol vertical profile. Calculations with this type of aerosol had shown the better agreement with observational data than the continental and maritime ones. Last fact can be clearly visible in the Figure 3, where the observational values of polarization ratio mixed derivative (see Appendix A) is compared with model calculations for different aerosol types. Diverse of degree of polarization between measurements and calculations for urban aerosol does not exceed several percents for all observed wavelengths.

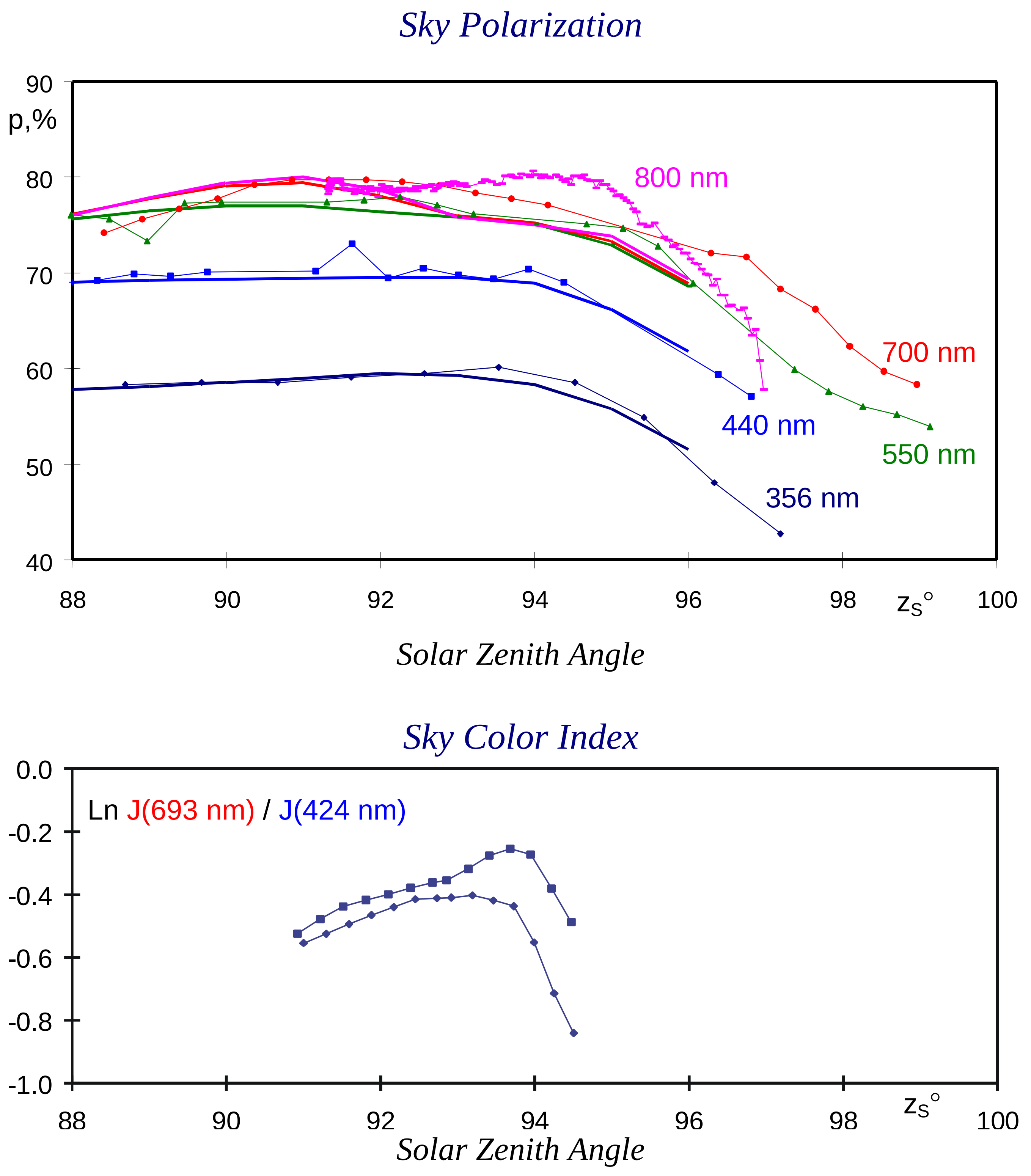

**Figure 2 (a).** The dependencies of twilight sky polarization, p, at the zenith on the solar zenith distance, $z_S$, for different wavelengths: 356 nm (the evening, July 31th, 1997), 440 nm (the morning, July 28th, 2000), 550 nm and 700 nm (the evening, August 4th, 2000), 800 nm (the evening, December 17th, 1999) compared with calculated curves for the urban type of aerosol (bold solid lines). **(b).** The dependencies of twilight sky color index on the solar zenith distance for two morning observations of July, 1994 [8].

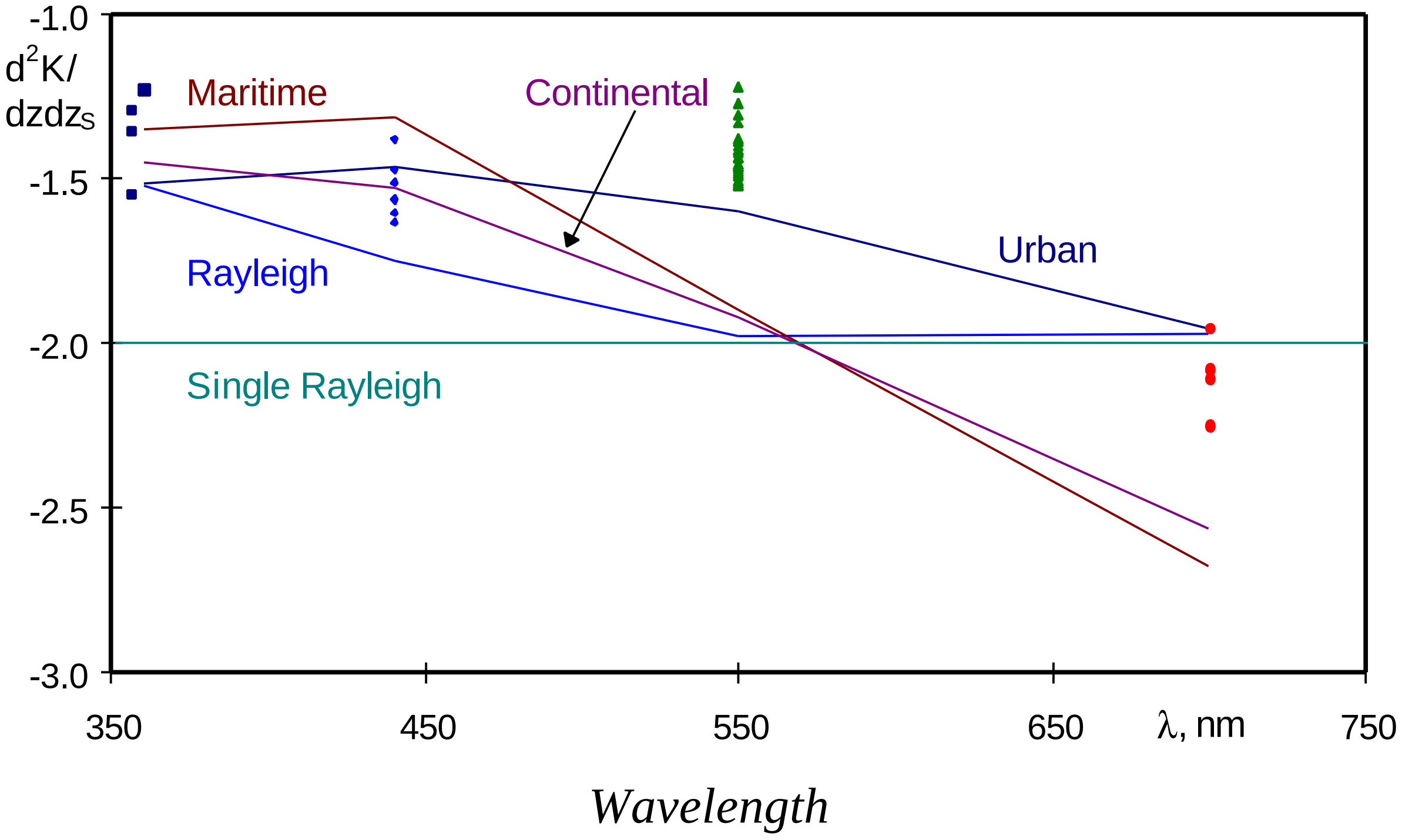

**Figure 3.** The value of polarization ratio mixed derivative obtained from 1997, 2000 and 2002 observations (dots) compared with radiative transfer simulation for pure Rayleigh scattering and different types of aerosol (solid lines). Solar zenith angle is equal to 90°.

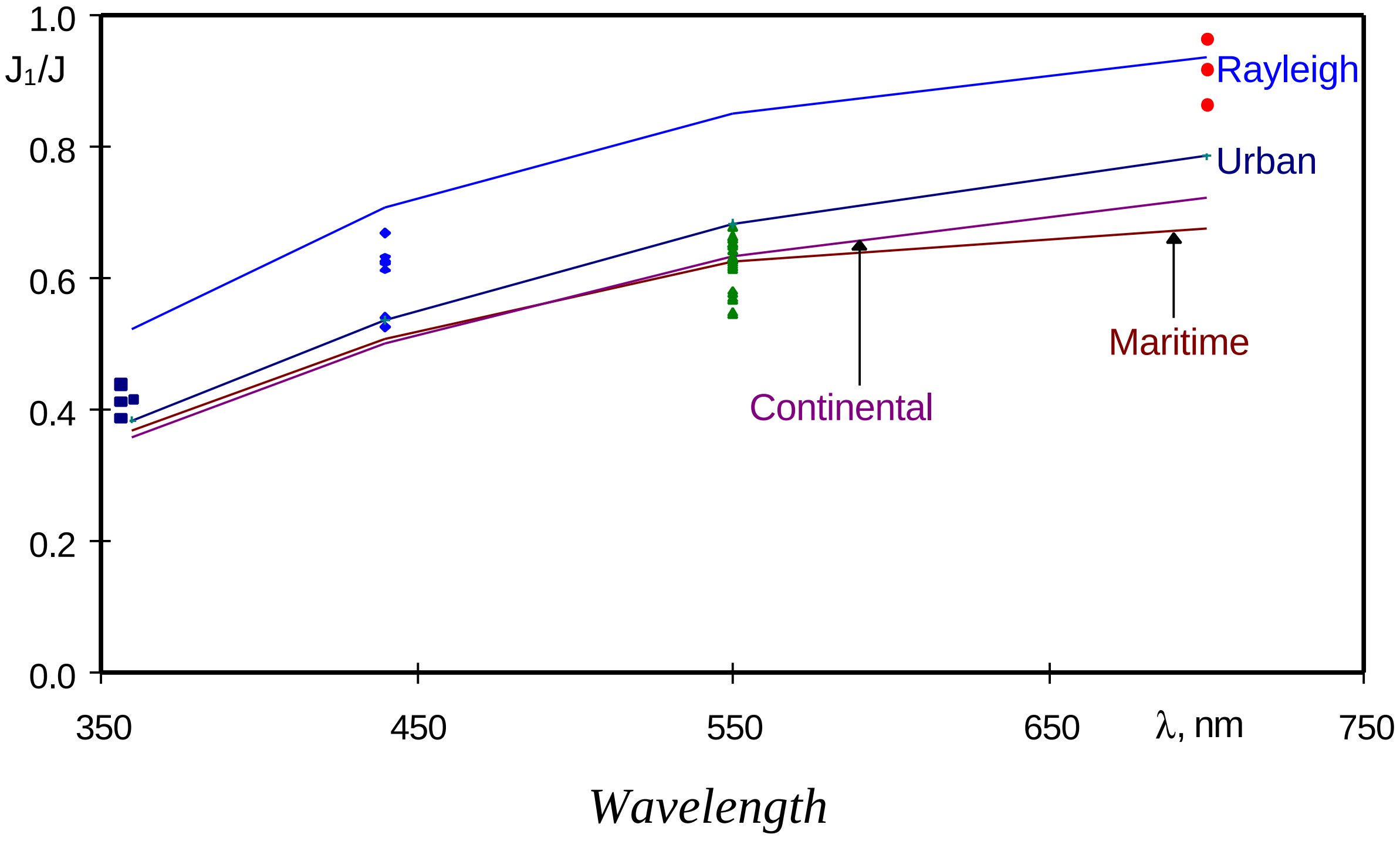

**Figure 4.** Estimation of the single scattered light contribution at the zenith by semi-empirical model for 1997, 2000 and 2002 observations (dots) compared with radiative transfer simulation for pure Rayleigh scattering and different types of aerosol (solid lines). Solar zenith angle is equal to 90°.

Both measurements and calculations show that the degree of polarization is strongest in the red part of spectrum. For shorter wavelengths the polarization is weaker, but its dependence on the solar zenith depth remains the same. This quantity is almost constant or even slightly rises from the sunset to the depressions about 4-5 degrees and then starts to fall. Here we should notice that the sky color has the same behavior: getting slightly red before, it turns blue at the same depressions of Sun, as it can be seen in Figure 2b which is based on the results of [8]. At the same solar zenith distance the maximum polarization point is returning to the zenith (see Figure 1). As it was shown in [2], the nature of color and polarization changes of the sky at this twilight stage is common and related with single- to multiple-scattering ratio changes.

## 6.2. Contribution of single scattered light

Figure 4 shows the single scattering contribution for the sunrise (sunset) moment at the zenith for observations in 1997, 2000 and 2002 calculated by the semi-empirical model in comparison with the radiative transfer calculations for pure Rayleigh scattering and different aerosol models. The single scattered light (compared with multiple scattered one) has an excess in long-wave part of spectrum and stronger polarization, and it explains the observed correlation of twilight polarization and color evolution which is visible in Figure 2 (a and b). This comparison confirms that estimations of single scattering by this semi-empirical model [1] have higher accuracy than earlier works, which underestimated multiple scattering.

## 6.3. Vertical profile retrieval

We carried out numerical experiments on aerosol profile retrieval using one wavelength 800 nm and one (zenith) direction of observations. Such measurements can not provide simultaneous determination of the concentration and microphysical characteristics of aerosol. Therefore, microphysical characteristics of aerosol were taken as known. The model [7] was used: continental aerosol below 10 km, and stratospheric aerosol above 10 km. For future convenience of comparison of retrieved data with data of lidar sounding, the profiles of aerosol backscattering ratios was retrieved. The algorithm [4] of retrieval is based on optimal statistical estimation method. Measurement errors were taken as typical for observation at Kislovodsk in 2002. A simulation was performed in approximation of single scattering, thus, correction of observed polarized radiance is necessary if the algorithm is applied to real data.

The dependence of the retrieval error on the vertical resolution is given in Figure 5. The errors were calculated for a measurement at 60 solar zenith angles from $z = 90.1°$ to $z = 96°$ with step 0.1°. Percentages were calculated for background aerosol profile. Notice that in periods subsequent to powerful volcanic eruptions the aerosol concentration is enhanced. In such periods, the percentage error may become less than 10%. At heights below 5 km, the relative error decreases due to high aerosol concentrations in the atmospheric surface layer (not shown). As the vertical resolution degrades to 5 km, the retrieval error decreases below 40% for all heights above 11 km. At a further degradation of the vertical resolution, the retrieval error does not decrease sufficiently. Figures 6–8 illustrate the numerical experiment on aerosol profile retrieval. The true aerosol layer is characterized by thickness of 2 km and range of aerosol concentration variation equal to the background concentration. At a resolution of 1 km, it is difficult to reveal such a model layer against the noise background; at a resolution of 2 km, the layer manifests itself rather well; and, at a resolution of 3 km, the layer is clearly pronounced but its form is noticeably distorted.

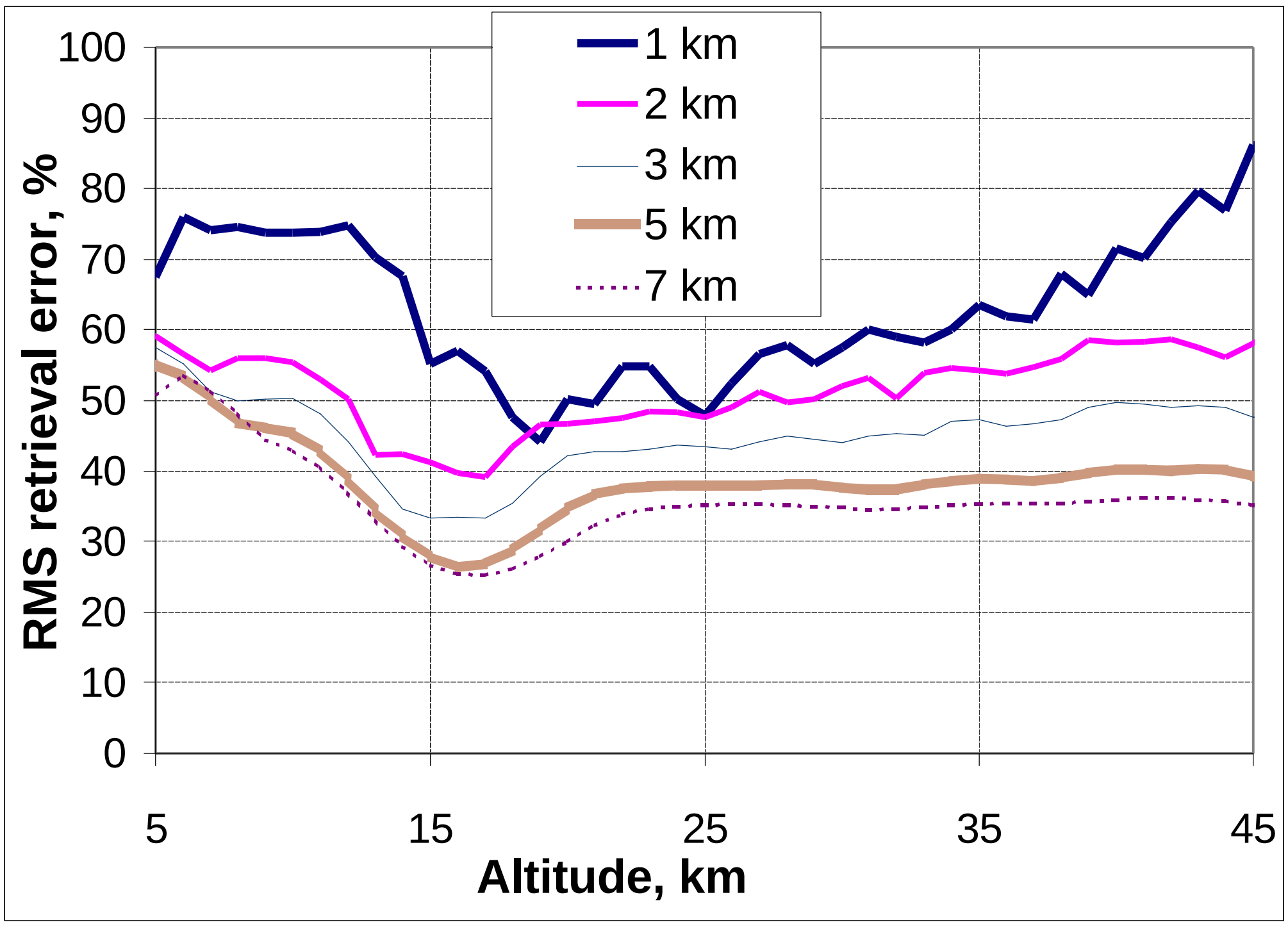

**Figure 5.** Percentage rms retrieval errors of aerosol concentration for different values of vertical resolution.

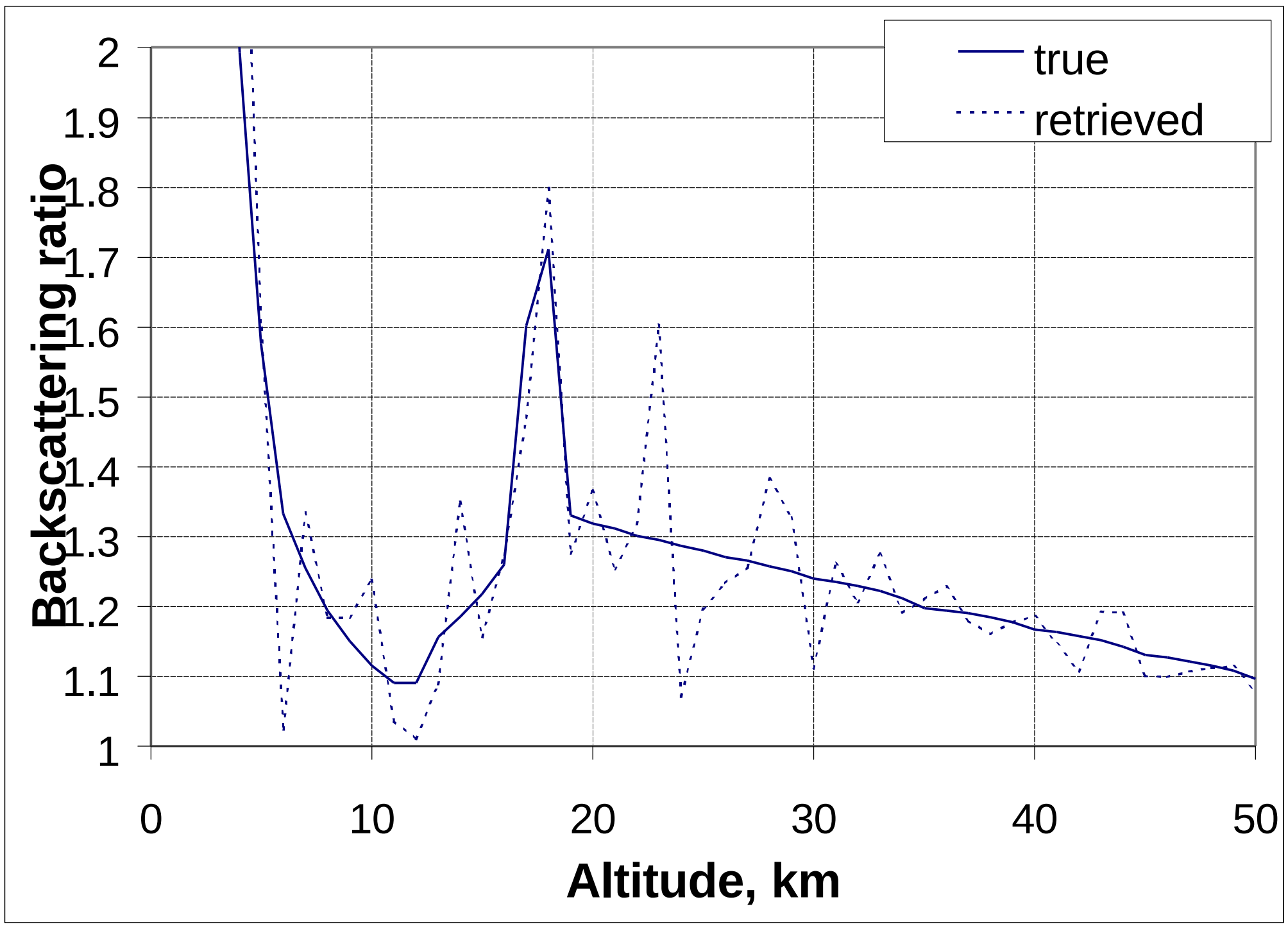

**Figure 6.** Numerical experiment on retrieval of aerosol backscatterring ratio with resolution equal to 1 km.

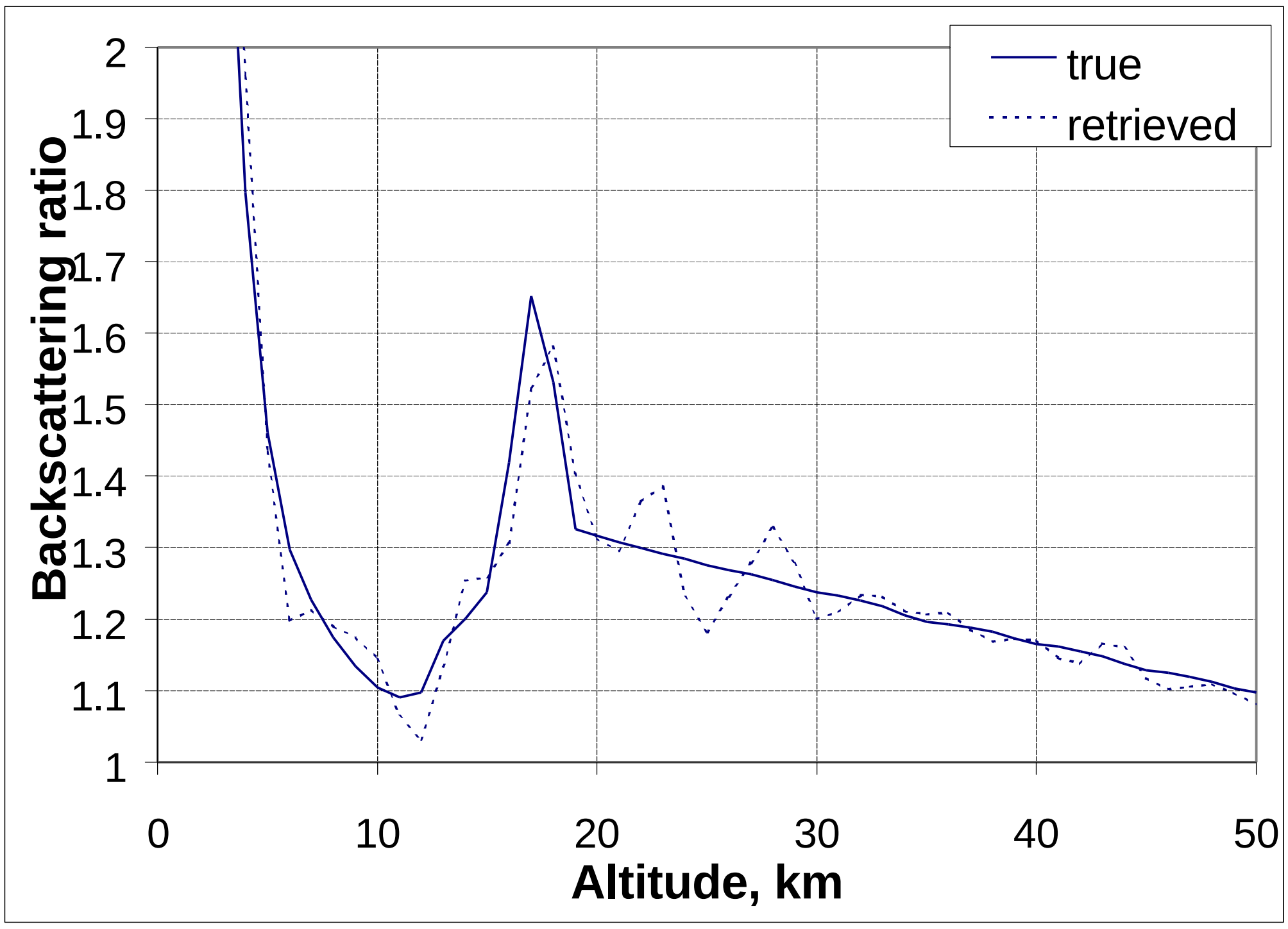

**Figure 7.** Numerical experiment on retrieval of aerosol backscatterring ratio with resolution equal to 2 km.

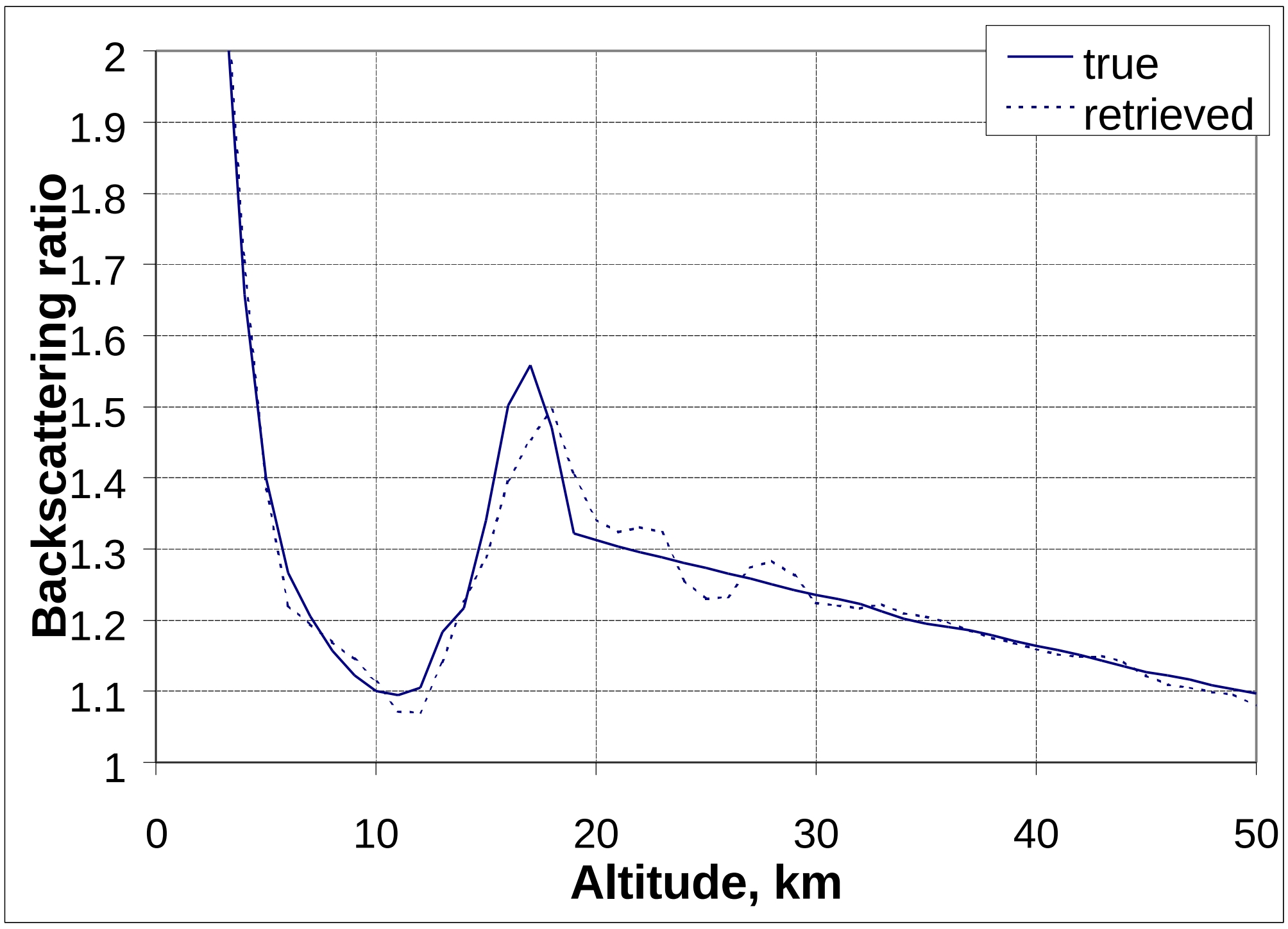

**Figure 8.** Numerical experiment on retrieval of aerosol backscatterring ratio with resolution equal to 3 km.

## 7. CONCLUSIONS

Observed during 1997-2002 at the different locations the degree of polarization well agrees with radiative transfer calculations for the urban type of microphysics [7] and background profile of aerosol. Coincidence of measurements and calculations is within 6% for the most solar zenith angles and angles of observation.

The percentage of single scattered light at the zenith during twilight is equal to about 40% for 360 nm, 60% for 440, 70% for 550 nm and reaches 80% for 700 nm, what confirmed by both the semi-empirical model based on the measurements, and the radiative transfer calculations. This fact is the reason of the lower sky polarization at shorter wavelengths. These percentages remain almost constant until the sun's depth under horizon 4-5°, when the part of single scattering rapidly decreases, and the sky polarization falls. At the sun's depth about 10° single scattered light completely disappears on the multiple scattered background.

A theoretical error analysis shows that measurements carried out at 800 nm with the IAP instrument has accuracy enough to detect a 2-km aerosol layer, which exceeds background concentration by more than 50% at 12-25 km or by 60% at 5-45 km, or detect a 5-km aerosol layer, which exceeds background concentration by more than 25% at 18-20 km or by 40% at 11-45 km. Notice that in periods of the enhanced aerosol concentration (after powerful volcanic eruptions) this retrieval error may become less than 10%.

## Appendix A. Definitions.

The polarization degree $p=(J_{\perp}-J_{||})/(J_{\perp}+J_{||})$, where $J_{\perp(||)}$ is the sky brightness value for the polarization plane perpendicular (parallel) to the scattering plane.

The polarization ratio $K=J_{||}/J_{\perp}=(1-p)/(1+p)$.

The polarization ratio mixed derivative $d^2K/dzdz_S$ ($rad^{-2}$), where $z$ and $z_S$ are the zenith distances of observed point and the Sun, respectively. The value is calculating at $z=0$ and $z_S=90°=\pi/2$.

The twilight sky color index $\ln J(\lambda_1)/\ln J(\lambda_2)$, where $\lambda_1$ and $\lambda_2$ are two different wavelengths ($\lambda_1>\lambda_2$).

## Appendix B. Optical model of atmosphere.

Rayleigh scattering: Rayleigh extinction profile for 550 nm is taken from [9]. Wavelength dependence is $\lambda^{-4}$. Depolarization factor is ignored.

Ozone: Model of G. Krueger with total ozone content 345 DU with MODTRAN cross sections.

Lambertian surface albedo is equal to 0.5.

## Appendix C. Aerosol microphysics defined in WMO 1986

**Table 1. Kinds of aerosol particles defined in WMO 1986.**

| Kind of particle | Dust | Water-soluble | Oceanic | Soot |
|---|---|---|---|---|
| Function of size distribution | lognormal | lognormal | lognormal | lognormal |
| Parameters | 0.5 mkm, 2.99 | 0.005 mkm, 2.99 | 0.3 mkm, 2.51 | 0.0118 mkm, 2.00 |

**Table 2. Percentage of aerosol particle in different kinds of aerosol defined in WMO 1986.**

| Kind of aerosol | Dust | Water-soluble | Oceanic | Soot |
|---|---|---|---|---|
| Urban | 17% | 61% | 0% | 22% |
| Continental | 70% | 29% | 0% | 1% |
| Maritime | 0% | 5% | 95% | 0% |

## REFERENCES

1. *Ugolnikov, O.S.* Twilight Sky Photometry and Polarimetry. The Problem of Multiple Scattering at the Twilight Time // *Cosmic Research*. 1999. V.37. P.159.
2. *Ugolnikov O.S., Maslov, I.A.* Multi-Color Polarimetry of the Twilight Sky. The Role of Multiple Scattering as the Function of Wavelength // *Cosmic Research*. 2002. V.40. P.224; also available as e-print physics/0112090 at www.arxiv.org.
3. *Postylyakov, O.V., Elansky, N.F., Elohov, A.S., Masleev, A.N., Orlov, M.N., and Sitnov, S.A.* Observations of polarized zenith-sky radiances during twilight with application to aerosol profile evaluation *IRS 2000 // Current Problems in Atmospheric Radiation*. A. Deepak Publishing, Hampton, Virginia. 2001. P.1197.
4. *Postylyakov, O.V.* Radiative transfer model MCC++ with evaluation of weighting functions in spherical atmosphere for usage in retrieval algorithms. // *Submitted to Adv. Space Res*. 2003.
5. *Postylyakov, O.V., Belikov, Yu.E., Nikolaishvili, Sh.S., Rozanov, A.* A comparison of radiation transfer algorithms for modeling of the zenith sky radiance observations used for determination of stratospheric trace gases and aerosol *IRS 2000 // Current Problems in Atmospheric Radiation*. A. Deepak Publishing, Hampton, Virginia. 2001. P.885.
6. *Postylyakov, O.V.* Linearized radiative transfer model MCC++ // *Abstract EAE03-A-12620*. (A poster presentation in Session AS28, Poster Area Esplanade P0864, on display on Tuesday, 8 April 2003).
7. WMO, World Climate Program: A preliminary cloudless standard atmosphere for radiation computation // *WCP-112, Radiation Commission, Int. Assoc. of Meteorol. and Atmos. Phys*. 1986.
8. *Ugolnikov, O.S.* Determination of Parameters of Light Scattering in the Earth's Atmosphere from Photometric Twilight Sky Observations // *Cosmic Research*. 1998. V.36. P.430.
9. *Zuev, V.V., Krekov, G.M.* Optical models of atmosphere. Leningrad, Gidrometeoizdat. 1986 (in Russian).